\documentclass[artical]{IEEEtran}

\usepackage{amsmath}

\ifCLASSINFOpdf
\usepackage[pdftex]{graphicx}
  
\else
 
\fi

\begin{document}
%
\title{Abel Dynamics of Titanium Dioxide Memristor Based on Nonlinear Ionic Drift Model}

\author{Weiran Cai,
        Frank Ellinger,~\IEEEmembership{Senior Member}, Ronald Tetzlaff,~\IEEEmembership{Member}, 
        and Torsten Schmidt
\thanks{Weiran Cai and Frank Ellinger are with the Chair for Circuit Design and Network Theory, and Ronald Tetzlaff and Torsten Schmidt are with the Chair for Fundamentals of Electronics, Fakult\"at Elektrotechnik und Informationstechnik, Technische Universit\"at Dresden, Helmholtzstrasse 18, Barkhausenbau, 01069 Dresden, Germany. e-mail: weiran.cai@tu-dresden.de.}
}

\maketitle

\begin{abstract}
We give analytical solutions to the titanium dioxide memristor with arbitary order of window functions, which assumes a nonlinear ionic drift model. As the achieved solution, the characteristic curve of state is demonstrated to be a useful tool for determining the operation point, waveform and saturation level. By using this characterizing tool, it is revealed that the same input signal can output completely different $u-i$ orbital dynamics under different initial conditions, which is the uniqueness of memristors. The approach can be regarded as an analogy to using the characteristic curve for the BJT or MOS transisitors. Based on this model, we further propose a class of analytically solvable class of memristive systems that conform to Abel Differential Equations. The equations of state (EOS) of the titanium dioxide memristor based on both linear and nonlinear ionic drift models are typical integrable examples, which can be categorized into this Abel memristor class. This large family of Abel memristive systems offers a frame for obtaining and analyzing the solutions in the closed form, which facilitate their characterization at a more deterministic level. 
\end{abstract}


\IEEEpeerreviewmaketitle

\section{Introduction}
\IEEEPARstart{M}{athematical} tools have been inspiring and pedestalling the development in fundamental electronics, in parallel with experimental means, ever since the birth of this scientific branch. Memristors were first postulated theoretically by L. O. Chua in 1971 ingeniously on the observation of formal symmetries in the mathematical links among four elementary electronic quantities, and was proved on the ground of electromagnetics\cite{Chua1}. The concept was revived by the realization of the titanium dioxide memristor in 2008\cite{Stru}, which has been responded by massive studies. Consequently, the mathematical structure of passive electronic elements has been extended to a large family consisting of memristive, memcapacitive and meminductive systems\cite{Chua2}\cite{Vent}. And recently, the class of heterogeneous memristive systems has further broadened the concept from the prevailing homogeneous systems\cite{Cai}. Despite of the richness in the structures, the understanding of memristive systems is still far less accurate and systematic, as compared to the existing theories of BJT or MOS transistors. This is mainly caused by the lack of proficient tools in solving the governing nonlinear differential equations. 

In fact, since the emergence of the titanium dioxide memristor, the dynamics of various systems have been mainly studied by numerical simulations. This is due to the rarity of integrable models (models that are solvable in the closed form) in memrisitve systems, unlike the cases of the various long-existing transistors, governed by algebraic equations. Numerical simulations have the merits of flexibility and versatility, but also suffer from the drawback that the physical relations of dominating factors to the dynamics can only be determined empirically or remain undetermined. In contrast,  solutions in the closed form can offer the insight in such relations and render the behaviors reasonable and predictable. In this paper, we give the analytical solution to the titanium dioxide memristor with arbitary order of window functions, which takes into account the nonlinear ionic drift model. We demonstrate that the characteristic curve of state, as the achieved solution, is a useful tool in determining the operation point, waveform and saturation level, similar to the characteristic curve for BJT or MOS transisitors. At the meantime, a natural question is proposed whether there can be some classes of memristor models in general that are solvable or conditionally solvable in closed form. In Ref.\cite{Geor} and \cite{Drak}, the authors heading for this question pioneered in proposing a class of memristive systems that conform to Bernoulli Differential Equations and argued that all current- or voltage-controlled memristors are classified therein. However, due to the implicit dependence of the memristance $\mathcal{M}(q(t))$ ($\mathcal{M}(\phi(t))$) on the current (voltage), this argument is only true for very special cases, \textit{e.g.}, the titanium dioxide memristor based on linear ionic drift model. We propose another class of (conditionally) solvable memristive systems that conform to (generalized) Abel Differential Equations. The equations of state of the titanium dioxide memristor based on both linear and nonlinear ionic drift models are typical integrable examples, which can be categorized into this Abel memristor class.

\section{Analytical Solutions to the Nonlinear Ionic Drift Model}
The titanium dioxide memristor, based on the mechanism of tunable doped region, was first implemented by B. Strukov, \textit{et al}, and was formulated with a linear ionic drift model\cite{Stru}. However, the linear model fails in revealing the boundary effects, which should differ the rate of the change in the size of the doped region in the bulk from that at the boundary. To overcome this serious drawback, Y. N. Joglekar, \textit{et al}, proposed a nonlinear ionic drift model in \cite{Jogl}, which modifies the linear drift equation with a window function of defined orders $p$:
\begin{equation}\label{f}
F(x)=1-(2x-1)^{2p}
\end{equation}
with $p\in \mathcal{N}$. It characterizes a milder drop in the rate of the dopant scale $w$ at the boundaries with a smaller $p$, while a sharper drop with a larger $p$. The entire set of equations describing the dynamics therefore reads as
\begin{subequations}\label{tm}
    \begin{alignat}{2}
     &v(t)=\left[ R_{ON}\frac{w(t)} D + R_{OFF}\left(1-\frac{w(t)} D\right)\right] i(t)  \label{tm1}\\
     &\frac{dw(t)}{dt}=\frac{\mu_V R_{ON}} D i(t) F\left(\frac{w(t)} D\right)  \label{tm2}
    \end{alignat}
\end{subequations}
This is an analytically solvable model. By eliminating $i(t)$ in Eq. (\ref{tm}), the equation of the state-variable $w(t)$ can be written as 
\small\begin{eqnarray}\label{wt}
\left[w(t)+\frac{R_{OFF}}{R_{ON}-R_{OFF}} D\right] \frac{dw(t)}{dt}
=  \frac{\mu_V R_{ON} v(t)}{R_{ON}-R_{OFF}} F\left(\frac{w(t)} D\right) \nonumber
\end{eqnarray}\normalsize
\begin{equation}\label{wt_num}
~
\end{equation}
We name it as the \textit{Equation of State} (EOS) of the memristor. In fact, the EOS corresponding to different values of $p$ can be categorized into a class of Abel Differential Equations\cite{Poly}. When $p=0,1$, Eq. (\ref{wt}) is an Abel equation of the second kind, which conforms to the equation
\begin{equation}\label{abel2}
\left[y(t)+\eta(t)\right]\frac{dy(t)} {dt}=f_2(t)y(t)^2+f_1(t)y(t)+f_0(t)
\end{equation}
When $p\ge 2$, it is a generalized Abel equation in the form of 
\begin{eqnarray}\label{abel2m}
\left[y(t)+\eta(t)\right]\frac{dy(t)} {dt}&=&f_{M-1}(t)y(t)^{M-1}+\nonumber\\
&&\cdot\cdot\cdot+f_2(t)y(t)^2+f_1(t)y(t)+f_0(t)\nonumber\\
\end{eqnarray}
There have been solutions to vast forms of the Abel equations and well-established qualitative analysis tools. However, one can observe that the Abel equation in Eq. (\ref{wt}) has a special form, which can be integrated directly by variable seperation. Making use of the special integral (for $m \in \mathcal{N}$) in the hypergeometric form
\begin{equation}\label{int}
\int \frac 1 {1-z^m}dz=z\cdot _2\mathcal{F}_1\left(\left[1,\frac 1 m\right],\left[1+\frac 1 m\right],z^m\right)
\end{equation}
where $_2\mathcal{F}_1$ is the Gaussian hypergeometric function (also known as the (2, 1)-ordered generalized hypergeometric function)\cite{Abra}, the general solution to Eq. (\ref{wt}) for any $p\in \mathcal{N}$ can be given in the following implicit form:
\small\begin{eqnarray}\label{gs}
&&(R_{ON}+R_{OFF}) y(t) \cdot_2\mathcal{F}_1\left(\left[1,\frac 1 {2p}\right],\left[1+\frac 1 {2p}\right], y(t)^{2p}\right)\nonumber\\
&+&\frac 1 2 (R_{ON}-R_{OFF})  y(t)^2\cdot_2\mathcal{F}_1\left(\left[1,\frac 1 {p}\right],\left[1+\frac 1 {p}\right], y(t)^{2p}\right)\nonumber\\
&=&\mu_V R_{ON} \Phi(t)+C(y(0))
\end{eqnarray}\normalsize
where $\Phi(t)$ denotes the integral of the input voltage signal (the flux) $\Phi(t)\equiv\int_0^t v(\tau)d\tau$, $y(t)$ is a rescaled variable $y(t)\equiv1-2w(t)/D\in [-1,1]$ and $C(y(0))$ is the integration constant determined by the initial condition $y(0)$ (or $w(0)$). Equivalently, the solution can be also expressed in the form of series
\small
\begin{equation} \label{ser}
\begin{split}
&(R_{ON}+R_{OFF})\sum_{k=0}^\infty \frac {y(t)^{2pk+1}}{2pk+1}\\
&+\frac 1 2 (R_{ON}-R_{OFF}) \sum_{k=0}^\infty \frac {y(t)^{2(pk+1)}}{pk+1}
=\mu_V R_{ON} \Phi(t)+C(y(0))
\end{split}
\end{equation} 
\normalsize
by using the relation between the hypergeometric function and gamma functions, and noting the property $\Gamma (z+1)=z\Gamma (z)$:
\begin{eqnarray}\label{2f1}
_2\mathcal{F}_1\left([1,a],[1+a],z\right) &=& \sum_{k=0}^\infty z^k \frac{\Gamma (a+1)\Gamma (a+k)}{\Gamma (a)\Gamma (a+k+1)}\nonumber\\
&=& \sum_{k=0}^\infty \frac a {a+k} z^k
\end{eqnarray}
In spite of the complexity in the general form, the analytical solutions to the system with window functions to the lowest orders can be given in a simpler form as follows, noting the Taylor series of logarithmic function and of inverse trigonometric function $arctan(x)$:\\
$p=0:$ (with no window function)
\begin{eqnarray}\label{p0}
&&\frac 1 2 (R_{ON}-R_{OFF})w(t)^2+ R_{OFF}Dw(t)\nonumber\\
&=&\mu_V R_{ON}\Phi(t)+C(w(0)) 
\end{eqnarray}
$p=1:$
\begin{eqnarray}\label{p1}
&&\frac 1 4 D^2 \left(R_{OFF}\ln \left|\frac{w(t)} D\right|-R_{ON}\ln\left|1-\frac{w(t)} D\right|\right)  \nonumber\\
&=&\mu_V R_{ON} \Phi(t)+C(w(0))  
\end{eqnarray}
$p=2:$
\begin{eqnarray}\label{p2}
&&\frac {\Delta} {16} \ln \left|\frac{2w(t)^2}{D^2}-\frac{2w(t)} D+1\right| + \frac {\bar{\Delta} } 8\arctan \left(\frac{2w(t)} D-1\right) \nonumber\\
&+& \frac {D^2} 8  \left( R_{OFF}\ln \left|\frac{w(t)} D\right|-R_{ON}\ln\left| 1-\frac{w(t)} D\right| \right) \nonumber\\
&=&\mu_V R_{ON}\Phi(t)+C(w(0)) 
\end{eqnarray}
with 
\begin{eqnarray}\label{sym}
\Delta &\equiv& D^2 \left( R_{ON}-R_{OFF}\right)\nonumber\\
 \bar{\Delta} &\equiv&  D^2\left( R_{ON}+R_{OFF}\right)\nonumber
\end{eqnarray}
where $C(w(0))$ is the integration constant. The Abel dynamics of $w(t)$ is the essential to the system, which links the voltage, current, flux and charge:
\begin{eqnarray}\label{link}
q(t) \longleftrightarrow i(t)\longleftrightarrow w(t) \longleftrightarrow \Phi(t) \longleftrightarrow u(t). \nonumber
\end{eqnarray}

\section{Characteristic Curve for Determining Orbital Shapes}
\begin{figure}[!t]
\centering
\includegraphics[width=8.3cm]{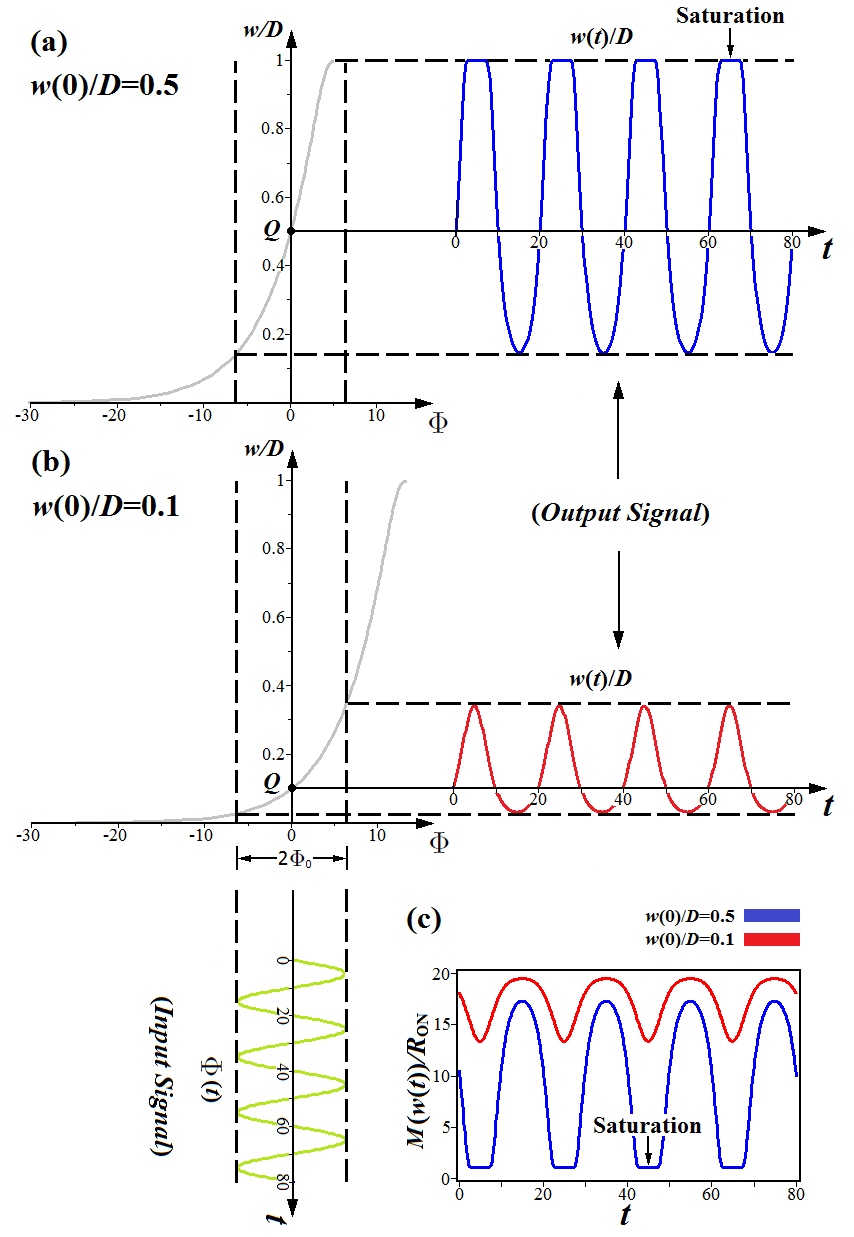}\\
\includegraphics[width=8.0cm]{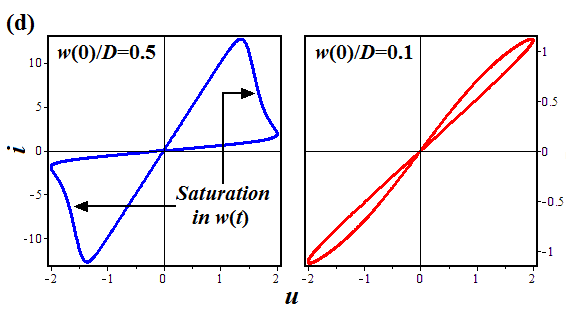}
\caption{Using the characteristic curve $\Phi-w$ for characterizing the TiO$_2$ memristor of nonlinear ionic drift model. (a) and (b) demonstrate the determination of the operation point and the amplitude of the state-variable with the input signal and the initial conditions, under $w(0)/D=0.5$ and $0.1$, respectively (suppose $p=1$). Q denotes the operation point, corresponding to the initial condition. (c) temporal variation in the memristance. (d) the $u-i$ phase dynamics. The orbital shapes are determined by the waveform of $w(t)$ in (a) and (b), including the saturations.  
}\label{fig1} 
\end{figure}

\begin{figure}[!t]
\centering
\includegraphics[width=7.2cm]{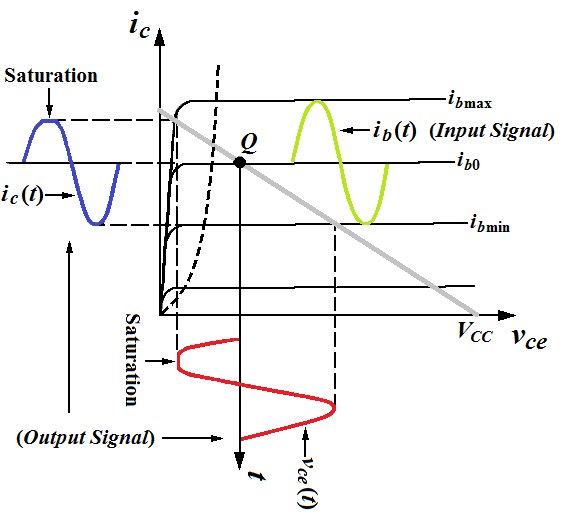}
\caption{Characteristic curve of transistors (BJT) as a tool for determining operation point, waveform and saturation level. The input current $i_b(t)$ is transfered to the output current $i_c(t)$ and voltage $v_{ce}(t)$ with saturation at the peaks in the illustrated case. Q denotes the operation point along the load line.
}\label{fig2} 
\end{figure}

Although the titanium dioxide memristor has been a focus of study in the area of memristive systems for long time, its dynamical behaviors still cannot been clearly characterized in a deterministic manner, compared to that for transistors. Numerical simulations, though flexible and versatile, are not proficient to reveal the inner relation between the parameters and the phase-planar dynamics. Critical questions, such as how the initial condition plays a role in the dynamics, which is crucial to a nonlinear system, and what determines exactly the deviation of the $u-i$ orbital shape from a standard hyteretic loop, remain unexplained. The $\Phi-w$ curve in the closed form, as given in Eq. (\ref{ser}), which we name as the \textit{Characteristic Curve of State}, provides a useful tool for characterizing the dynamics of the memristor, which can determine the operation point, amplitude, waveform and saturation level, as using the characteristic curve in analyzing BJT or MOS transistors. The procedure is explained as follows:

Suppose that the input voltage signal is $v(t)=v_0\cos(\omega t)$, which corresponds to a flux band with the width of $2\Phi_0= 2v_0/\omega$. For memristors, two factors are needed to determine the dynamical behavior of the nonlinear system: the input signal and the initial condition. We can see that the initial condition for the titanium oxide memristor actually functions as an operation point. The analytical solution in Eq. (\ref{gs}) indicates that different initial conditions $C(w(0))$ are represented as a translation of the characteristic curve $w(\Phi)$, which determines the intersection of the characteristic curve and the input flux band. When the steeper part of the curve is within the interval $[-\Phi_0,+\Phi_0]$, the input signal $\Phi(t)$ is mapped onto the dopant scale $w(t)$ with a larger amplitude; and vice versa. The amplitude of $w(t)$ determines directly the change in the memristance $\mathcal{M}(w(t))$ (Fig. \ref{fig1}(c)) and hence the shape of the $u-i$ orbital shape (Fig. \ref{fig1}(d)). Similar to the case in transistors, the characteristic curve also reveals the saturation level in $w(t)$, as it occurs, for example, in the upper peaks ($w\rightarrow D$) in Fig. (\ref{fig1}(a)). In fact, the saturation in $w(t)$ will result in a deformation of the $u-i$ orbital shape (see the triangular shaped orbit with a sharper negative resistance in Fig. \ref{fig1}(d) for $w(0)/D=0.5$), as compared with a standard-shaped hysteretic loop (see Fig. \ref{fig1}(d) for $w(0)/D=0.1$). It is of special interest that the same input signal can output completely different $u-i$ orbital dynamics under different initial conditions, which is the uniqueness of memristors!

The narrowing of the $u-i$ orbital shape is usually observable when raising the operation frequency\cite{Stru}. However, the above analysis implies that this phenomenon can also be obtained by varying the initial condition. It is preceivable that a sufficiently large input signal will cause saturation on both upper and lower peaks ($w\rightarrow 0$ and $D$) in $w(t)$, which will produce a heavier deformation in the $u-i$ orbit. The drastic difference in orbital shapes reminds us that to control the initial condition properly is critical in the correct operation of memristors, which is highly related to the reliability of memristive networks. 

In fact, this characterizing method can be well regarded as an analogy of using the characteristic curve in transistors for similar purposes (see Fig. \ref{fig2}), in which the output signal $i_c(t)$ or $v_{ce}(t)$ encounters saturation when the operation point $Q$ on the load line is close to the saturation region of the transistor. Therefore, with the characteristic curve $\Phi-w$ in the closed form, one is able to determine the main characteristics of the dynamics of the memristor prior to amounts of numerical simulations in the same manner as for transistors.

\section{Abel Dynamics in Memristive Systems}  
\begin{figure}[!t]
\centering 
\includegraphics[width=7.5cm]{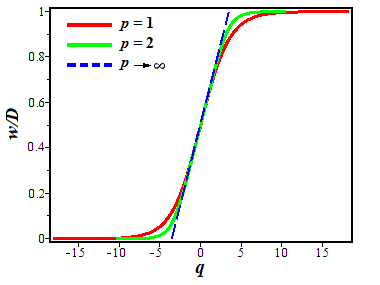}
\caption{The relation between the charge $q$ and the dopant scale $w$. The $p\rightarrow\infty$ limit approaches the linear relation without window functions.
}\label{fig3} 
\end{figure}

As researchers have seen that analytical solutions can offer a deterministic characterizing tool for memristive systems, a natural question is raised whether there can be certain classes of analytically solvable memristive systems. This problem was first addressed by P. Georgious, \textit{et al.} In Ref. \cite{Geor} and \cite{Drak}, an argument was made that any current-controlled memristor(similar for the case of voltage-controlled memristor) belongs to the class of Bernoulli memristive systems. More specifically, the nonlinear first order ordinary differential equation of the memristor on the current $i(t)$:
\begin{equation}\label{Mqphi}
\frac{di(t)}{dt}-\frac{dv(t)}{dt} \frac{i(t)}{v(t)} = -\frac{d}{dt}[\mathcal{M}(q(t))] \frac{i^2(t)}{v(t)}
\end{equation}
is argued to conform to a Bernouli Differential Equation, in the general form of 
\begin{equation}\label{B}
\frac{dy(t)}{dt}+f(t)y(t) = g(t)y(t)^n
\end{equation}
where $\mathcal{M}(q(t))$ denotes the current-controlled memristance. As demonstrated in the referred papers\cite{Geor}\cite{Drak}, this classification is correct for the titanium dioxide memristor based on linear ionic drift model as a current-controlled memristor\cite{Stru}. However, this argument is not valid in general, due to the implicit dependence of the memristance $\mathcal{M}(q(t))$ ($\mathcal{M}(\phi(t))$) on the current (voltage). The titanium dioxide memristor model with nonlinear dopant drift, discussed in the last section, is a typical system beyond this category, despite of its current-controlled memristive nature. To see this point, it is noticeable that the relation between the charge $q$ and the dopant scale $w$ is modified by a hypergeometric factor for $p>0$:
\begin{equation}\label{qw}
\begin{split}
&q(w)=\frac{D^2}{2\mu_V R_{ON}} \left(\frac{2w} D-1\right)\\
&\times _2\mathcal{F}_1\left(\left[1,\frac 1 {2p}\right],\left[1+\frac 1 {2p}\right],\left(\frac{2w} D-1\right)^{2p}\right)+B,
\end{split}
\end{equation}
where $B$ is an integration constant. Noting the relation in Eq. (\ref{2f1}), $q(w)$ is a series of odd-ordered terms of $\left(\frac{2w} D-1\right)$ and hence a bijective mapping between $w$ and $q$ (see Fig. \ref{fig3}). Therefore, the memristance $M(w)$ can be expressed by $M[w(q)]$, which indicates that the system with window functions remains a current-controlled memristor. However, due to the relation of the term
\begin{equation}\label{Mqw}
\begin{split}
\frac{d}{dt}[\mathcal{M}(q(t))]&=\frac{d\mathcal{M}(q(w))}{dw}\left(\frac{dq(w)}{dw}\right)^{-1}\frac{dq}{dt}\\
&=\frac{R_{ON}-R_{OFF}}{D}\left(\frac{dq(w)}{dw}\right)^{-1}i(t),
\end{split}
\end{equation}
in which $dq(w)/dw$ is a function of $w(t)$ by Eq. (\ref{qw}), the right-handside of Eq. (\ref{Mqphi}) has a dependence on the unsolved $w(t)$. Henceforth, the $u-i$ behavior of the system does not conform to a Bernoulli equation. 

Instead of Bernoulli dynamics, the EOS of the titanium dioxide memristor based on both linear and nonlinear ionic drift models conform to the Abel Differential Equations, as demonstrated in the previous section. This categorization can be generalized to a large family of memristive systems--the Abel memristor class, which offers a large number of analytical solutions. Let us consider a memristive system with single state-variable. Suppose that $v(t)$ is the input signal, the system equations are possible to be reduced to a differential equation on the state variable, by eliminating the current $i(t)$. Assuming a general memristive system\cite{Vent}
 \begin{subequations}\label{msys}
    \begin{alignat}{2}
    v(t)&=g(x,i,t)\cdot i \label{msys_a}\\
\dot x&= f(x,i,t)  \label{msys_b}
    \end{alignat}
  \end{subequations}
can be reduced to a differential equation on the state-variable $x(t)$ or more generally a bijective mapping of it $y[x(t)]$, we define it as the general \textit{Equation of State} (EOS) of the system:
\begin{equation}\label{msys2}
\mathcal{S}\left(y(t),\dot y(t),v(t),t\right)=0
\end{equation}
This reduction is possible for most memristive systems, since the governing equations are algebraic equations of $i(t)$. If the EOS has one of the following polynomial forms
\begin{equation}\label{abel_1}
\frac{dy(t)} {dt}=f_3(t)y(t)^3+f_2(t)y(t)^2+f_1(t)y(t)+f_0(t)
\end{equation}
and
\begin{equation}\label{abel_2}
\left[y(t)+\eta(t)\right]\frac{dy(t)} {dt}=f_2(t)y(t)^2+f_1(t)y(t)+f_0(t)
\end{equation}
namely, the Abel Differential Equations of the first kind and of the second kind, we classify such systems into Abel memristive systems of the first kind and of the second kind, respectively. Actually, the Abel memristive systems of the second kind can be converted to the first kind under the transformation $y(t)+\eta(t)=1/Y(t)$; inversely, an Abel system of the first kind can also be converted to an Abel system of the second kind under the transformation $y\rightarrow y^{-1}$\cite{Poly}.  

The Abel class is an important family in studying memristive systems for its formal commonty and solvability. Although the Abel equations are not integrable for arbitary coefficient functions, there have been solutions to a large family of specific forms, especially solutions to periodically driven systems. A standard procedure in solving an Abel equation of the second kind is to make a reduction to the canonical form of $\Omega\cdot\Omega_z-\Omega=R(z)$, with $\Omega_z$ being the derivative to the transformed variable z, to which a large number of solutions have been provided in Ref. \cite{Poly}. On the other hand, various tools for qualitative analysis have been established in analyzing the solutions and the stability\cite{Alwa}\cite{Alva}. 

The two classes of memristive systems can be further enlarged to the generalized Abel memristive systems with polynomials to higher orders
\begin{equation}\label{abel_1m}
\frac{dy(t)} {dt}=f_M(t)y(t)^M+\cdot\cdot\cdot+f_2(t)y(t)^2+f_1(t)y(t)+f_0(t)
\end{equation}
or its equivalent form
\begin{eqnarray}\label{abel_2m}
\left[y(t)+\eta(t)\right]\frac{dy(t)} {dt}&=&f_{M-1}(t)y(t)^{M-1}+\nonumber\\
&&\cdot\cdot\cdot+f_2(t)y(t)^2+f_1(t)y(t)+f_0(t)\nonumber\\
\end{eqnarray}
It is observable that when $f_k(t)=0$, for $k=0, 2,...,(M-1)$, Eq. (\ref{abel_1m}) is again reduced to a Bernoulli Differential Equation\cite{Poly} (this is, however, an equation on the state-variable, instead of on the current or voltage). As supported by the achievements in the theory of ordinary differential equations, the generalized Abel class provides an important family of memristive systems in the polynomial form, under which a large number of analytical solutions are obtainable to the correponding models.

\section{Conclusions}
We have given analytical solutions to the titanium dioxide memristor based on the nonlinear ionic drift model. The achieved characteristic curve of state as the solution has been demonstrated to be a useful tool in characterizing the memristor, similar to the characteristic curve for BJT or MOS transisitors. By using this characterizing tool, we have discovered that the same input signal can output completely different $u-i$ orbital dynamics under different initial conditions, which is the uniqueness of memristors! Based on this model, we have further proposed the Abel class of analytically solvable memristive systems. The EOS of the titanium dioxide memristor based on both linear and nonlinear ionic drift models are typical integrable examples, able to be categorized into this Abel memristor class. This large family of Abel memristive systems has offered a frame for characterization of the systems at a more deterministic level via the solutions in the closed form.

\ifCLASSOPTIONcaptionsoff
  \newpage
\fi




\end{document}